\documentstyle[prd,aps]{revtex} 
\begin{document}
\draft
\title{Comment on the  Herzlich's proof of the Penrose inequality}

\date{\today}
\author{Edward Malec}
\address{Jagiellonian University,
Institute of Physics, {3}0-59 Krak\'ow, Reymonta 4, Poland}
\author{Krzysztof Roszkowski}
\address{Jagiellonian University,
Institute of Physics, {3}0-59 Krak\'ow, Reymonta 4, Poland}
\maketitle
\begin{abstract}
Recently Herzlich proved a Penrose-like inequality with a coefficient
being a kind of a Sobolev constant. We show  that this constant
tends to zero for charged black holes approaching
maximal Reissner -  Nordstr\"om   solutions.  The
method proposed by Herzlich is not appropriate for  charged matter
with nonzero global charge.
\end{abstract}
\pacs{ }

One of the most interesting problems of General Relativity was formulated  in
1973 by R. Penrose \cite{Pen} in his attempt to specify conditions under
which a cosmic censorship hypothesis can be broken.

Specified entirely in terms of initial data,
the {\bf Penrose Conjecture} says the following.

{\it Let(M,g) be a 3-dimensional asymptotically flat
Riemannian manifold with a compact, connected, minimal and stable (inner)
boundary $\partial$M which is a topological 2-sphere. Suppose also that the scalar
curvature of (M,g) is nonnegative. Then its mass, if defined, satisfies:}
$$
m\geq \sqrt{\frac{Area(\partial M)}{16\pi}}
$$
{\it and equality is achieved if and only if (M,g) is a spacelike Schwarzschild
metric.}

A  proof of this inequality is  known only in special cases (\cite{gibbons},
\cite{malec94}). In 1997 Marc Herzlich presented
a refreshingly new approach to that
problem in the case of  momentarily static initial data.
His  main theorem cited below in extenso \cite{Herz} shows a
Penrose - like inequality with a functional coefficient
$\tau = {2\sigma \over 1+
\sigma }$.

{\it Let(M,g) be a 3-dimensional asymptotically flat
Riemannian manifold with a compact, connected, minimal and stable (inner)
boundary $\partial$M which is a topological 2-sphere. Suppose also that
the scalar
curvature of (M,g) is nonnegative. Then its mass, if defined, satisfies:}
$$
m\geq \frac{2\sigma}{1+\sigma}\sqrt{\frac{Area(\partial M)}{16\pi}},
$$
{\it where $\sigma$ is a dimensionless quantity defined as}
$$
\sigma=\sqrt{\frac{Area(\partial M)}{\pi}}\inf_{\phi\in
C^{\infty}_c,\phi\neq0}\frac{||d\phi||^2_{L^2(M)}}
{||\phi||^2_{L^2(\partial M)}}.
$$
{\it Moreover, equality is achieved if and only if (M,g) is a spacelike
Schwarzschild metric of mass} $\frac{1}{4}\sqrt{Area(\partial M)/\pi}$.

In what follows we show that $\tau $ can be arbitrarily small in the
case of spherical  charged black holes that are almost maximal.

Let us assume a spherically symmetric spacetime
and choose isotropic coordinates.
Then spatial part of the line element reads
\begin{equation}
ds^2=f^4(r) (dr^2+ r^2 d\Omega^2),
\end{equation}
where $ d\Omega^2 $ is the line element on a unit sphere
and $0\le r < \infty $.

It is convenient to normalize the function $\phi$   so that
 $\phi $ is equal to 1 at
the inner boundary. Restriction to the set of spherically symmetric
minimizers yields
\begin{eqnarray}
&&\sigma=2\inf_{\phi :\phi\left(r_0\right)=1}
\frac{\int^{\infty}_{r_0}\left(
\partial_r\phi\right)^2f^2r^2dr}{r_0f^2\left( r_0\right) }.
\label{3}
\end{eqnarray}
A minimizing funtion $\phi $  satisfies a  harmonic equation
which reads, assuming spherical symmetry,
\begin{equation}
{1\over r^2f^6}
{d\over dr}(r^2f^2\phi')=0.
\end{equation}
Therefore $\phi ' = {C\over r^2 f^2}$ and
\begin{equation}
\phi(r)=1+\int^r_{r_0}\frac{C}{f^2(r')r'^2}dr'.
\label{4}
\end{equation}
The condition that   $\phi $ vanishes at infinity
leads to  the integration constant $C$
\begin{equation}
C={-1\over \int_{r_0}^{\infty }ds {1\over s^2f^2(s)}}.
\label{5}
\end{equation}
The integrand of  (\ref{3}) can be written as
\begin{eqnarray}
&&\int^{\infty}_{r_0}\phi '\left(\phi 'f^2r^2\right)dr=
\phi\left(\phi 'f^2r^2\right)|^{\infty}_{r_0}-
\int^{\infty}_{r_0}\phi '\frac{d}{dr}\left(\phi 'f^2r^2\right)dr=
\nonumber \\
&&-\bigl( \phi\left(\phi 'f^2r^2\right)\bigr) (r_0)=|C|.
\label{6}
\end{eqnarray}
From this we  finally arrive at
\begin{equation}
\sigma =\frac{2}{r_0f^2\left(r_0\right)
 \int_{r_0}^{\infty }ds {1\over s^2f^2(s)}}.
\label{7}
\end{equation}
Assume   a fixed momentarily static Cauchy slice.
One can smoothly join an interior geometry of a charged matter region
with the    external (electrovacuum)
geometry as given by the conformal factor f,
\begin{equation}
f=\sqrt{1+\frac{m}{r}+\frac{m^2-q^2}{4r^2}}.
\label{8}
\end{equation}
Explicit integration  of the integral of (\ref{5}) yields,
in electrovacuum,
\begin{equation}
\int_{r_0}^{\infty }ds {1\over s^2f^2(s)} ={1\over |q|}\ln (1+
{|q|\over r_0 +m/2-|q|/2});
\label{9}
\end{equation}
therefore
\begin{equation}
\sigma = {2|q|\over
\ln (1+{|q|\over r_0 +m/2-|q|/2} ) (r_0+m+\frac{m^2-q^2}{4r_0})}.
\label{10}
\end{equation}
If  {\it $r_0$} is a coordinate radius of a minimal surface,
then
$$
\frac{d}{dr}\left(r^2f^4\right)=0
$$
In electrovacuum  the last equation yields the radius $r_0$ of a minimal
sphere as a function of $m$ and $q$,
\begin{equation}
r_0=\sqrt{\frac{m^2-q^2}{4}}.
\label{11}
\end{equation}
That implies  that at the specified radius
\begin{equation}
f\left(r_0\right)=
\sqrt{2+\frac{2m}{\sqrt{m^2-q^2}}}.
\end{equation}
We may choose parameters so that $m^2-q^2$ is arbitrarily small.
Then the denominator in (\ref{10}) becomes arbitrarily large
and  both $\sigma$ and $\tau $ may be  as small as one wishes  \cite{Tod}.
In this limit the result of Herzlich reduces merely to the positivity of the
asymptotic mass $m$. We would like to point out that this does not imply that
the Penrose inequality is not valid for the charged matter (in fact it holds
true), but  that the method discussed above is not appropriate.

It is interesting to note that charged matter  again and again  causes
troubles  in various conjectures.  Bonnor has shown \cite{bonnor}
that charged matter falsifies the standard form of the hoop conjecture
\cite{thorn}.
Then  a compactness criterion of Yau and Schoen \cite{yau}
for the formation of horizons has been shown to fail \cite{bizon}.
An isoperimetric inequality proposed by Gibbons
  \cite{gibbons} can be broken by distributions with a bulk
of charge outside of a horizon \cite{malec94}.

{\bf Acknowledgement}
One of the authors (EM) gratefully
acknowledges discussions with Marc Herzlich.
This paper is in part supported by the KBN grant   2 PO3B 090 08.


\begin{references}
\bibitem{Pen}R. Penrose, Naked Singularities Ann. N. Y. Acad. Sci. {\bf 224},
125-134 (1973)
\bibitem{gibbons} G. Gibbons, in {\it Global Riemannian Geometry}
ed. N. J. Willmore and N. J. Hitchin, Cambridge University Press (1984).
\bibitem{malec94} E. Malec and N. O' Murchadha, {\it Phys. Rev.}
{\bf D49}, 6931(1994); Appendix in Phys. Rev. {\bf D54},  4792(1996)
\bibitem{Herz} M. Herzlich, {Commun. Math. Phys.} {\bf 188}, 121-133 (1997).
\bibitem{Tod} Niall O'Murchadha and Marc
Herzlich informed us recently that
Paul Tod arrived independently at the same conclusion.
\bibitem{bonnor} W. B. Bonnor, {\it Phys.Lett.} {\bf A99}, 424(1983)
\bibitem{thorn} C. W. Misner, K. Thorne, J. Wheeler, Gravitation (Freeman,
San Francisco 1973), p.867.
\bibitem{yau}
R. Schoen, S. T. Yau, {\it Commun.Math.Phys.} {\bf 90}, 575(1983)
\bibitem{bizon}  P. Bizo\'n, E> Malec and N. O'Murchadha,
{\it Class. Quantum  Grav.} {\bf 6}, 961(1989).
\end{references}
\end{document}